\documentclass[journal,a4paper,twocolumn,oneside]{IEEEtran}
\usepackage{graphicx,amssymb,amsmath,amsthm,cite,url}
\pdfoutput=1

\newtheorem{theorem}{Theorem}

\def\mathlette#1#2{{\mathchoice{\mbox{#1$\displaystyle #2$}}%
                               {\mbox{#1$\textstyle #2$}}%
                               {\mbox{#1$\scriptstyle #2$}}%
                               {\mbox{#1$\scriptscriptstyle #2$}}}}
\renewcommand{\vec}[1]{\mathlette{\boldmath}{#1}}

\newcommand{\prob}[1]{\Pr\left[#1\right]}

\DeclareMathOperator*{\sign}{sign}

\newcommand{\figwidth}{0.95\columnwidth}
\newcommand{\figwidthB}{0.6\columnwidth}

\begin{document}

\title{Analysis and Design of Binary\\Message-Passing Decoders}

\author{	
Gottfried Lechner,
Troels Pedersen, and
Gerhard Kramer
\thanks{Manuscript submitted to the \emph{IEEE Transactions on Communications}, April 2010.

Gottfried Lechner is with the Institute for Telecommunications Research, University of South Australia (Email: gottfried.lechner@unisa.edu.au). Troels Pedersen is with the Department of Electronic Systems, Aalborg University, Denmark (Email: troels@es.aau.dk). Gerhard Kramer was with Bell Labs, Alcatel-Lucent, Murray Hill, NJ. He is now with the University of Southern California, Los Angeles, CA (Email: gkramer@usc.edu).

Parts of this work have been presented at the IEEE International Symposium on Information Theory (ISIT) 2007 and at the Australian Communications Theory Workshop (AusCTW) 2010.
}%
}

\maketitle

\begin{abstract}
Binary message-passing decoders for low-density parity-check (LDPC) codes are studied by using extrinsic information transfer (EXIT) charts. The channel delivers hard or soft decisions and the variable node decoder performs all computations in the L-value domain. A hard decision channel results in the well-know Gallager B algorithm, and increasing the output alphabet from hard decisions to two bits yields a gain of more than 1.0 dB in the required signal to noise ratio when using optimized codes. The code optimization requires adapting the mixing property of EXIT functions to the case of binary message-passing decoders. Finally, it is shown that errors on cycles consisting only of degree two and three variable nodes cannot be corrected and a necessary and sufficient condition for the existence of a cycle-free subgraph is derived.
\end{abstract}


\section{Introduction}
\label{sec:intro}
Gallager introduced low-density parity-check (LDPC) codes \cite{Gallager1962,Gallager1963} and also presented message-passing decoding algorithms that exchange only {\em binary} messages between the variable and check nodes. These algorithms are referred to as Gallager A and Gallager B \cite{Richardson2001} depending on how the variable-to-check node messages are computed. The algorithms have small memory requirements and low complexity implementations, especially of the check node decoder, and they have found practical use in high-speed applications, e.g. optical transmission systems \cite{ITU-T2004}. However, the complexity advantages come at the cost of a significant loss in performance.

In this work, we use extrinsic information transfer (EXIT) charts~\cite{Gallager1962,Gallager1963,tenBrink1999,Ashikhmin2004} to analyze and design binary message-passing algorithms. Interestingly, in contrast to non-binary message passing where they are an approximation, EXIT charts are exact for the case of binary messages (and infinite-length code ensembles) since the mutual information describes the probability densities of the messages precisely. Furthermore, the EXIT functions for binary message-passing algorithms can be derived analytically \cite{Gallager1962,Gallager1963}.

Binary message-passing algorithms were studied in \cite{Ardakani2005,Ardakani2004a} where the authors showed that optimum algorithms must satisfy certain symmetry and isotropy conditions. In contrast to majority based decision rules, we assume that the variable node decoder converts all incoming messages to L-values~\cite{Hagenauer1996}, performs decoding in the L-value domain and applies a hard decision on the result. Note that for these algorithms, there always exist majority decision rules that can be derived in a straightforward way as shown in Section~\ref{sec:majority}. This general approach assures that the symmetry and isotropy conditions are satisfied and the algorithms can be extended for systems where the channel provides more information than hard decisions, while the variable and check node decoder still exchange binary messages only. This reduces the gap between optimum decoding and binary message-passing decoding, while the complexity is kept low.

Our main contributions are as follows:
\begin{itemize}
\item We derive a framework that allows binary message-passing algorithms to incorporate various quantization schemes for the channel messages. Increasing the number of quantization bits from one to two leads to a significantly improved decoding performance.
\item We identify certain structures of the factor graph which cannot be corrected by binary message-passing decoders. Therefore, in addition to the stability condition, the degree distribution has to satisfy another constraint in order to avoid error floors. An important consequence is that regular LDPC codes with variable node degree three cannot be decoded without an error floor.
\end{itemize}

The rest of this paper is organized as follows. In Section~\ref{sec:preliminaries}, we introduce basics and definitions which are used in Section~\ref{sec:exit} to derive the EXIT functions of the variable and check node decoders. In Section~\ref{sec:irregular}, we show how the EXIT functions can be used to optimize the code and derive constraints on the degree distributions. Section~\ref{sec:implementation} considers practical aspects of binary message-passing decoders and simulation results are presented in Section~\ref{sec:simulations}.

\section{Preliminaries}
\label{sec:preliminaries}
For binary message-passing decoders, the extrinsic channel \cite{Ashikhmin2004} of the variable and check node decoder is represented as a binary symmetric channel (BSC) with input $X$ and output $Y$ both with alphabet $\{+1,-1\}$. Let $\epsilon$ denote the crossover probability of the BSC which we assume to be less than or equal to 0.5. Since there is a one-to-one relation between mutual information $I(X;Y)$ and crossover probability for the BSC, we can equivalently describe those channels using
\begin{align}
\label{equ:cap_bsc}
I(X;Y) = 1 - h_b(\epsilon)
\end{align}
where $h_b(\cdot)$ denotes the binary entropy function
\begin{align}
h_{b}(\epsilon) = -\epsilon \log_{2}(\epsilon)-(1-\epsilon) \log_{2}(1-\epsilon).
\end{align}
The variable node decoder converts all messages to L-values using
\begin{align}
L(y) & = \log\frac{\prob{X=+1|Y=y}}{\prob{X=-1|Y=y}}\nonumber\\ &= \log\frac{\prob{Y=y|X=+1}}{\prob{Y=y|X=-1}},
\end{align}
where $y$ is a realization of $Y$ and we assumed that $\prob{X=+1}=\prob{X=-1}=1/2$. Defining the \emph{reliability} associated with a BSC as
\begin{align}
\label{equ:reliability}
D = \log\frac{1-\epsilon}{\epsilon} \geq 0
\end{align}
allows to express the L-value as
\begin{align}
\label{equ:llrcomp}
L(y) = y \cdot D.
\end{align}
Throughout the paper, random variables are denoted by uppercase letters and their realizations are denoted by lowercase letters. The indices $v$, $c$, $a$, and $e$ stand for variable node decoder, check node decoder, a-priori and extrinsic, respectively.

\section{EXIT Functions of Component Decoders}
\label{sec:exit}

\subsection{Check Node Decoder}
A check node of degree $d_c$ of a binary message-passing algorithm computes the output message of each of its edges as the product of the other $d_c-1$ edge inputs when using the alphabet $\{+1,-1\}$. Let $\epsilon_{ac}$ denote the average bit error probability at the input of the check nodes. The corresponding a-priori crossover probability of the extrinsic channel is therefore also $\epsilon_{ac}$. We define $I_{ac}=1-h_b(\epsilon_{ac})$ so that $\epsilon_{ac}=h_b^{-1}(1-I_{ac})$ where $h_b^{-1}(x)$ takes on values in the interval $[0,1/2]$. The crossover probability at the check node output is \cite[Lemma 4.1]{Gallager1963}
\begin{align}
\label{equ:exit_check}
\epsilon_{ec} = f_c(\epsilon_{ac};d_c) = \frac{1 - (1 - 2\epsilon_{ac})^{d_c-1}}{2}
\end{align}
where $f_c$ is the EXIT function of a check node of degree $d_c$. Using (\ref{equ:exit_check}) and (\ref{equ:cap_bsc}), we define $I_{ec} = 1 - h_b(\epsilon_{ec})$. The inverse of the EXIT function in (\ref{equ:exit_check}) reads
\begin{align}
\label{equ:exit_check_inv}
\epsilon_{ac} = f_c^{-1}(\epsilon_{ec};d_c) = \frac{1 - \left(1 - 2\epsilon_{ec}\right)^\frac{1}{d_c-1}}{2}.
\end{align}

\subsection{Variable Node Decoder}
For a variable node of degree $d_v$, every outgoing message $L_{ev,j}$ along edge $j$ of the variable node is given by
\begin{equation}
\label{equ:varoperation}
L_{ev,j} = L_{ch} + \sum_{i=1;i \neq j}^{d_v}L_{av,i}, \hspace{1cm} j=1,\ldots,d_{v}
\end{equation}
where $L_{ch}$ is the L-value from the channel (see \eqref{equ:Lch} below) and $L_{av,i}$ is the L-value from the check nodes along edge $i$ of the variable node.
To perform this summation, all messages must be converted to L-values using (\ref{equ:reliability}) and (\ref{equ:llrcomp}), where we assume that the variable node decoder {\it knows} the parameters of the communication and extrinsic channels. We show in Section~\ref{sec:simulations} how the decoder can be implemented without this knowledge. In the following, we assume the communications channel is an additive white Gaussian noise channel with binary input (BIAWGN) and noise variance $\sigma_{n}^2$ so the received values $y$ are converted to L-values as
\begin{align}
\label{equ:Lch}
L_{ch} = \frac{2}{\sigma_{n}^2}y
\end{align}
before being quantized. Conditioned on $X=+1$, the unquantized L-values are Gaussian random variables with variance $\sigma^2_{ch}=4/\sigma_{n}^2$ and mean $\mu_{ch}=\sigma^2_{ch}/2$ \cite{Richardson2001a}. In the following, we derive the EXIT functions for three different quantization schemes.

\subsubsection{Hard Decision Channel}
\label{sec:hard}
Consider the case where the receiver performs hard decisions. Then the decoder's communication channel can be modeled as a BSC with crossover probability
\begin{align}
\epsilon_{ch} = \int_{-\infty}^{0} g(l) dl
\end{align}
where $g(l)=p_{L|X}(l|+1)$ is the conditional transition probability of the actual communication channel. Any symmetric channel with binary input is completely described by $g(l)$ which is therefore sufficient to analyze the decoding behavior. Let $D_{ch}$ and $D_{av}$ denote the reliabilities of the decoder's communication and extrinsic channels, respectively (see \cite[Fig. 2 and Fig. 3]{Ashikhmin2004}). The variable node decoder computes the outgoing message on edge $j$ by using \eqref{equ:varoperation} with $L_{ch}=y \cdot D_{ch}$. The outgoing message transmitted to the check node decoder is the sign of $L_{ev}$. To compute the error probability of the message, we consider two cases. First, assume that the channel message is in error. This error is corrected if the sum over the $L_{av,i}$ in \eqref{equ:varoperation} can overcome the incorrect sign on $L_{ch}$, i.e., if
\begin{align}
-D_{ch} - n_c D_{av} + (d_v-1-n_c) D_{av} \ge 0
\end{align}
where $n_c$ is the number of erroneous messages from the check nodes. An equivalent condition is $n_c \le t$ where
\begin{align}
\label{equ:tdefinition}
t = \left\lfloor\frac{D_{av}(d_v-1)-D_{ch}}{2D_{av}}\right\rfloor.
\end{align}
Similarly, if the channel message is correct, then $n_c$ has to be less than or equal to
\begin{align}
\label{equ:tbardefinition}
\bar{t} = \left\lfloor\frac{D_{av}(d_v-1)+D_{ch}}{2D_{av}}\right\rfloor
\end{align}
to result in a correct outgoing message.
Combining these two cases yields the error probability of the outgoing messages of the variable node decoder
\begin{eqnarray}
\label{equ:error_var_hard}
\epsilon_{ev} & = & f_v(\epsilon_{av};d_v,\epsilon_{ch})\nonumber\\
              & = & 1 - \epsilon_{ch} B\left(t; d_v-1, \epsilon_{av}\right)\nonumber\\
              &   & -(1-\epsilon_{ch}) B\left(\bar{t}; d_v-1, \epsilon_{av}\right)
\end{eqnarray}
where
\begin{equation}
B(k;n,p) = \sum_{i=0}^{k} \binom{n}{i} p^i (1-p)^{n-i}
\end{equation}
denotes the binomial cumulative distribution. The EXIT function \eqref{equ:error_var_hard} serves as a lower bound on EXIT functions for (appropriately designed) soft-decision detectors, see Section~\ref{sec:quantexamples} below.

\subsubsection{Soft Decision Channel}
\label{sec:soft}
In the limit of no quantization of the output of a BIAWGN channel, the crossover probability at the output of the variable node is
\begin{align}
\epsilon_{ev}
= & \prob{L_{ev,j} \le 0 | X=+1} \nonumber \\
= & \prob{\left. L_{ch} + \sum_{i=1;i\ne j}^{d_v} L_{av,i} \le 0 \right| X=+1} \nonumber \\
= & \sum_{z=0}^{d_v-1} \prob{N_c=z}\cdot\nonumber\\
& \prob{\left. -L_{ch} + \mu_{ch} \ge D_{av}(d_v-1-2z)+\mu_{ch} \right| X=+1} \nonumber \\
= & \sum_{z=0}^{d_v-1} b(z;d_v-1,\epsilon_{av}) Q\left(\frac{D_{av}(d_v-1-2z)+\mu_{ch}}{\sigma_{ch}}\right),
\label{equ:error_var_soft}
\end{align}
where $N_c$ is a random variable representing the number of erroneous messages from the check nodes, where
\begin{equation}
Q(\phi) = \frac{1}{\sqrt{2\pi}}\int_{\phi}^{\infty} e^{-\frac{\psi^2}{2}} d\psi
\end{equation}
is the familiar $Q$ function, and where
\begin{equation}
b(k;n,p) = \binom{n}{k} p^k (1-p)^{n-k}
\end{equation}
denotes the binomial probability mass function. The EXIT function \eqref{equ:error_var_soft} serves as an upper bound on the EXIT function of any quantization scheme, see Section~\ref{sec:quantexamples} below.

\subsubsection{Output Alphabets Larger than Binary}
\label{sec:quant}
\begin{figure}
  \centering
  \includegraphics[width=\figwidthB]{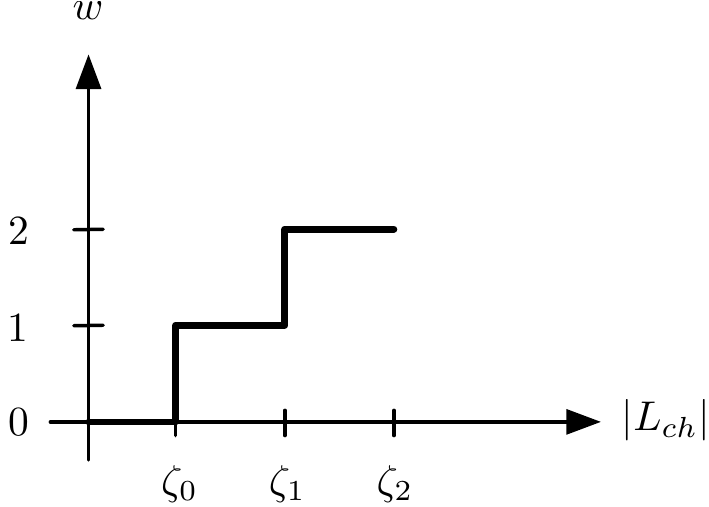}
  \caption{Quantization scheme for channel messages.}
  \label{fig:quantizer}
\end{figure}

Suppose a quantizer provides the sign $\sign(L_{ch})$ and a quantization index $w$, $w=0,\ldots,W$, of the magnitude $|L_{ch}|$. The boundary points of the quantizer are defined by the vector $\vec{\zeta}=[\zeta_0,\ldots,\zeta_W]$ where $0 \leq \zeta_0 < \zeta_1 < \dots < \zeta_W$. Such a quantization scheme is depicted in Figure~\ref{fig:quantizer}.

Following \cite{Land2005,Land2006}, this channel quantization scheme can be decomposed as $(W+1)$ BSCs.
Sub-channel $w$ is used with probability $p_w$, has cross-over probability $\epsilon_{ch,w}$ and reliability $D_{ch,w}$.
For $w>0$ we have
\begin{align}
p_w & = \int_{\zeta_{w-1}}^{\zeta_w} g(l) dl + \int_{-\zeta_w}^{-\zeta_{w-1}} g(l) dl, \label{eq:pk}\\
\epsilon_{ch,w} &= \frac{1}{p_w}\int_{-\zeta_w}^{-\zeta_{w-1}} g(l) dl, \hspace{0.5cm}\mbox{and}\\
D_{ch,w} &= \log\frac{1-\epsilon_{ch,w}}{\epsilon_{ch,w}}.  \label{eq:Dch}
\end{align}
We define sub-channel zero ($w=0$) as a BSC with crossover probability 0.5 \cite{Land2005,Land2006}. The parameters for sub-channel zero are
\begin{align}
\label{equ:p0}
p_0 &= \int_{-\zeta_0}^{\zeta_0} g(l) dl,\\
\epsilon_{ch,0} &= \frac{1}{2}, \hspace{0.5cm}\mbox{and}\\
D_{ch,0} &= 0.
\end{align}
The EXIT function of the overall channel is given by the expectation of the EXIT functions of the sub-channels \cite{Land2005,Land2006}
\begin{align}
\label{equ:exit_subchannels}
\epsilon_{ev} = \sum_{w=0}^W p_w \, \epsilon_{ev,w}
\end{align}
where $\epsilon_{ev,w}$ is given by \eqref{equ:error_var_hard} with $\epsilon_{ch}=\epsilon_{ch,w}$.

\subsection{Examples}
\label{sec:quantexamples}
\begin{figure}
  \centering
  \includegraphics[width=\figwidth]{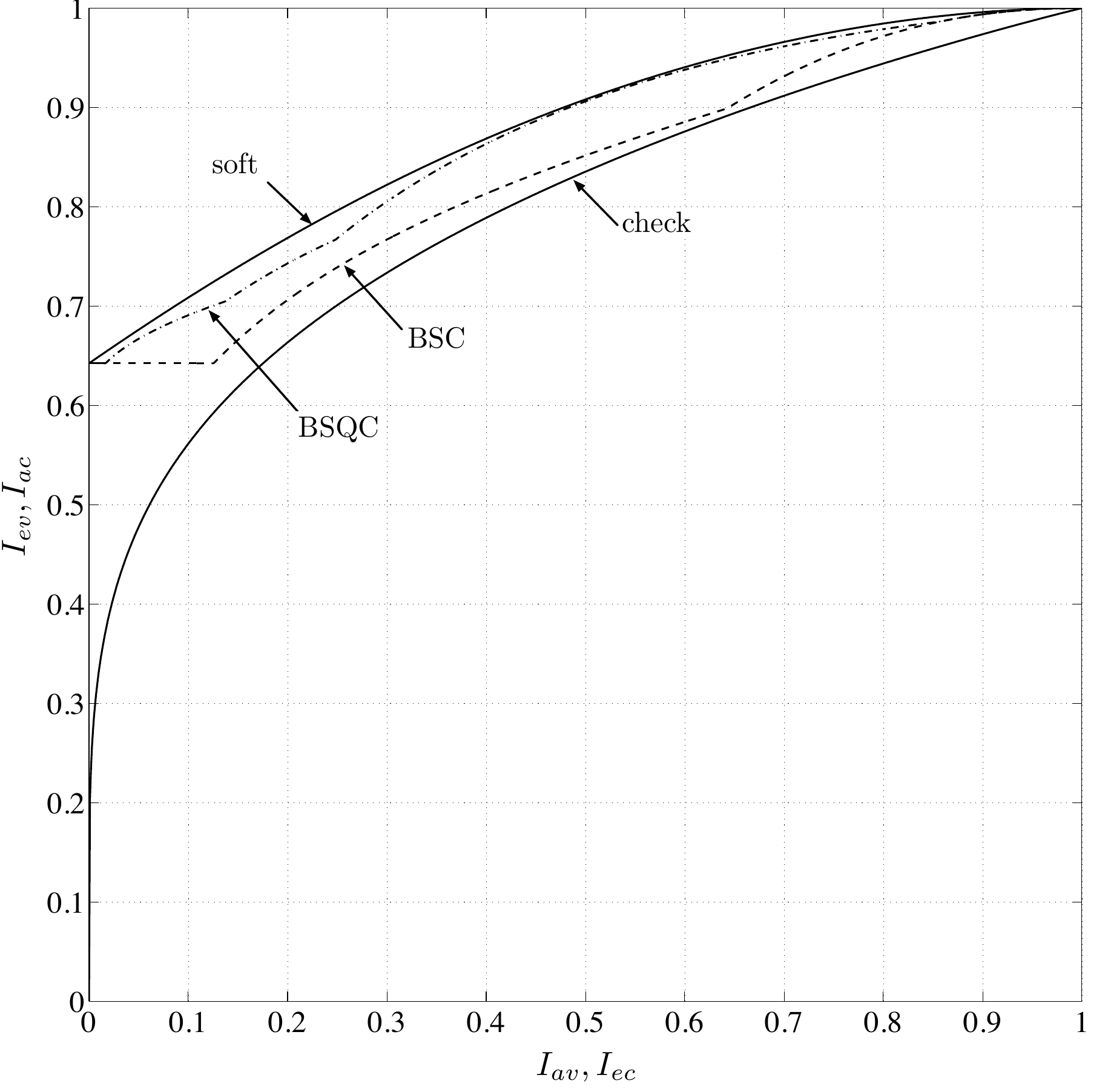}
  \caption{Binary message passing EXIT functions of check nodes with $d_c=6$ and variable nodes with $d_v=4$, $\zeta_1=1.90$ and $\sigma=0.67$ for BSC, BSQC and soft decision channel.}
  \label{fig:exit46}
\end{figure}

Figure~\ref{fig:exit46} shows examples of the EXIT functions for check and variable nodes for a regular LDPC code with variable node degree $d_{v}=4$ and check node degree $d_{c}=6$. In Figure~\ref{fig:exit46}, we use $I_{av}=1-h_{b}(\epsilon_{av})$, $I_{ev}=1-h_{b}(\epsilon_{ev})$ and similarly for $I_{ac}$ and $I_{ec}$. The EXIT function for the hard decision channel (BSC) changes its behavior at certain values of $I_{av}$. These values correspond to a change of the majority decision rule of the Gallager B algorithm \cite{Richardson2001}. The EXIT function for the soft decision channel is an upper bound for all quantization schemes. As an example for larger output alphabets, we consider a binary symmetric quaternary output channel (BSQC) where the output of the channel takes on values from $\{-D_{ch,2},-D_{ch,1},+D_{ch,1},+D_{ch,2}\}$. This channel output can be represented by a quantization using
\begin{align}
\vec{\zeta} =
\begin{bmatrix}
0, & \zeta_1, & \infty
\end{bmatrix}.
\end{align}
Notice that for this case \eqref{equ:p0} gives $p_{0}=0$.
In Figure~\ref{fig:exit46}, the EXIT function is shown for $\zeta_1=1.90$.

\section{Irregular Codes}
\label{sec:irregular}
In this section we consider irregular LDPC codes, their EXIT functions, and constraints that have to be satisfied to avoid error floors. The results will be used in Section~\ref{sec:simulations} to optimize the degree distribution of the variable nodes. We focus on check-regular codes, i.e. all check nodes have the same degree $d_{c}$, but the analysis extends in a simple way to check-irregular codes.

\subsection{Mixing of EXIT Functions}
\label{sec:mixing}
We first prove the following theorem that is stated in a different context in \cite{Tuechler2002}. Recall that $g(l)=p_{L|X}(l|+1)$ is the conditional probability density of a channel from $X$ to $L$.
\begin{theorem}
\label{th:mixing}
Consider a collection of channels $\{g_i(l)\}_i$ that satisfy $g_i(-l) = e^l g_i(l)$ for all $i$. We then have
\begin{align}
\label{equ:mixture}
\sum_i \lambda_i \; I(X_i ; L_i) = I(\overline{X};\overline{L})
\end{align}
where
\begin{align}
\label{equ:mixturepdf}
p_{\overline{L}\left|\overline{X}\right.}(l|+1)=\sum_i \lambda_i \; p_{L_i|X_i}(l|+1).
\end{align}
In other words, for $0\le\lambda_i\le1$ and $\sum_i\lambda_i=1$, the average EXIT function of the channel collection equals the EXIT function of the ``averaged" channel.
\end{theorem}
\begin{IEEEproof}
Since $g(l)=p_{L|X}(l|+1)=p_{L|X}(-l|-1)$, we obtain (note the integral limits)
\begin{align}
I(X;L)
= & h(L) - h(L|X) \nonumber \\
= & \int_{0}^{\infty} -\left[g(l)+g(-l)\right] \log \frac{g(l)+g(-l)}{2}\nonumber\\
  & \hspace{0.7cm}+ g(l) \log g(l) + g(-l) \log g(-l) dl.
\end{align}
Using the symmetry condition $g(-l)=e^{-l}g(l)$ this simplifies to
\begin{align}
I(X;L)
= & \int_{0}^{\infty} g(l)\left\{ e^{-l}\log\frac{2}{1+e^l} + \log\frac{2}{1+e^{-l}} \right\} dl
\end{align}
which is a linear operation on $g(l)$. Hence the order of summation and computation of mutual information may be swapped.
\end{IEEEproof}

Theorem~\ref{th:mixing} implies that the EXIT function of a mixture of codes can be computed as the average of the EXIT functions of the component codes as long as the channels to the L-values satisfy the property $g(-l) = e^l g(l)$. In the case of binary message-passing decoders, however, the outgoing L-values of the variable node are quantized to $\{+D_{ev},-D_{ev}\}$. This nonlinear operation prohibits the exchange of averaging and the computation of the mutual information. On the other hand, the mixture EXIT function can still be computed by averaging over the {\it crossover probabilities} instead of the EXIT functions \cite{Gallager1962,Gallager1963,Ardakani2002}, i.e., we have
\begin{align}
\label{equ:irregular}
\epsilon_{ev} & = \sum_{i} \lambda_{i}f_v(\epsilon_{av};i,\epsilon_{ch})
\end{align}
where $\lambda_{i}$ denotes the fraction of edges connected to a variable node of degree $i$ \cite{Richardson2001} and $f_{v}(\cdot)$ is given by \eqref{equ:error_var_hard}.
Furthermore, we can formulate the successful decoding constraint in terms of crossover probabilities
\begin{align}
\epsilon_{ev} = \sum_{i} \lambda_{i}f_v(\epsilon;i,\epsilon_{ch}) < f_c^{-1}(\epsilon)
\end{align}
for all $\epsilon \in (0,0.5)$. Expressing the design rate $R$ of a code as \cite{Richardson2001}
\begin{align}
R = 1 - \frac{\frac{1}{d_{c}}}{\sum_{i}\frac{\lambda_{i}}{i}}
\end{align}
leads to a linear program for maximizing the design rate:
\begin{align}
\label{equ:linprog}
\text{maximize} \hspace{5mm} & \sum_{i}\frac{\lambda_{i}}{i}\\
\text{subject to} \hspace{5mm} & \sum_{i} \lambda_{i}f_v(\epsilon;i,\epsilon_{ch}) < f_c^{-1}(\epsilon) \hspace{5mm}\forall\hspace{3mm}\epsilon \in (0,0.5)\nonumber\\
& \sum_{i}\lambda_{i}=1\nonumber\\
& 0\le\lambda_i\le1.\nonumber
\end{align}
In practice, the first constraint is evaluated for a fine grid of discrete $\epsilon$. This linear program enables the efficient optimization of the variable node degree distribution.

\subsection{Stability}
\label{sec:stability}
It has been shown in \cite{Richardson2001} that if the degree distribution of an LDPC code satisfies a stability condition, the decoder converges to zero error when starting from a sufficiently small error probability. The stability condition for the binary message-passing decoder is given in the following theorem.

\begin{theorem}
\label{th:stability}
An irregular LDPC code satisfies the stability condition under binary message-passing decoding (using quantized or unquantized channel messages) if and only if the variable node degree distribution satisfies
\begin{equation}
\label{equ:stability}
	\left(\lambda_{2} + 2\epsilon_{ch}\lambda_{3}\right)\left(d_{c}-1\right) < 1,
\end{equation}
where $\epsilon_{ch}$ denotes the error probability of a hard decision of the channel messages.
\end{theorem}
\begin{IEEEproof}
Let the superscript $^{(\ell)}$ denote the error probabilities at iteration $\ell$. Furthermore, let $\epsilon_{ac}^{(\ell)}=\epsilon_{ev}^{(\ell)}$ and $\epsilon_{av}^{(\ell+1)}=\epsilon_{ec}^{(\ell)}$ denote the error probabilities from the variable to check nodes and from the check to variable nodes, respectively. According to \eqref{equ:exit_subchannels}, the EXIT function of the variable node decoder with quantized channel messages is the expectation of the EXIT functions of the sub-channels. Combining \eqref{equ:exit_check}, \eqref{equ:error_var_hard} and \eqref{equ:exit_subchannels} leads to
\begin{align}
	\epsilon_{ev}^{(\ell+1)} & = \sum_{w=0}^{W} p_{w}f_{v}\left(\epsilon_{av}^{(\ell+1)} ; d_{v}, \epsilon_{ch,w}\right)\nonumber\\
& = \sum_{w=0}^{W} p_{w}f_{v}\left(f_{c}(\epsilon_{ev}^{(\ell)}; d_{c}) ; d_{v}, \epsilon_{ch,w}\right).
\end{align}
For small error rates, it suffices to consider only the first order Taylor series expansion over one iteration \cite{Richardson2001}. Stability implies that the error probability decreases over one iteration for sufficiently small error probabilities, i.e., we have
\begin{align}
\label{equ:oneiteration}
\lim_{\epsilon_{ev}^{(\ell)}\to 0}\frac{\partial\epsilon_{ev}^{(\ell+1)}}{\partial\epsilon_{ev}^{(\ell)}}\nonumber\\
& \hspace{-2cm}= \lim_{\epsilon_{ev}^{(\ell)}\to 0} \left(\sum_{w=0}^{W}p_{w}\,\left.\frac{\partial f_{v}(\eta;d_{v},\epsilon_{ch,w})}{\partial\eta}\right|_{\eta=f_{c}(\epsilon_{ev}^{(\ell)};d_{c})}\cdot\right.\nonumber\\
& \hspace{-0.5cm}\left. \left.\frac{\partial f_{c}(\eta;d_{c})}{\partial \eta}\right|_{\eta=\epsilon_{ev}^{(\ell)}}  \right)\nonumber\\
& \hspace{-2cm}= \lim_{\epsilon_{ev}^{(\ell)}\to 0} \left(\sum_{w=0}^{W}p_{w}\,\left.\frac{\partial f_{v}(\eta;d_{v},\epsilon_{ch,w})}{\partial\eta}\right|_{\eta=f_{c}(\epsilon_{ev}^{(\ell)};d_{c})}\right)\cdot \nonumber\\
& \hspace{-1.6cm}\lim_{\epsilon_{ev}^{(\ell)}\to 0} \left(\left.\frac{\partial f_{c}(\eta;d_{c})}{\partial \eta}\right|_{\eta=\epsilon_{ev}^{(\ell)}}  \right)\nonumber\\
& \hspace{-2cm}= \sum_{w=0}^{W}p_{w} \left(\lim_{\eta\to 0} \frac{\partial f_{v}(\eta;d_{v},\epsilon_{ch,w})}{\partial\eta}\right) \cdot \lim_{\eta\to 0} \left(\frac{\partial f_{c}(\eta;d_{c})}{\partial \eta}  \right) < 1,
\end{align}
where the final step follows because $f_{c}(\epsilon_{ev}^{(\ell)};d_{c})\to 0$ as $\epsilon_{ev}^{(\ell)} \to 0$.
For the check node \eqref{equ:exit_check}, we have
\begin{equation}
\label{equ:stabchk}
\lim_{\eta\to 0} \left(\frac{\partial f_{c}(\eta;d_{c})}{\partial \eta}  \right) = d_{c}-1.
\end{equation}
For the variable node decoder we start with the binomial cumulative distribution
\begin{align}
	\lim_{p \to 0} \frac{\partial}{\partial p}B(k;n,p) & = 	\lim_{p \to 0} \frac{\partial}{\partial p} \sum_{i=0}^{k} \binom{n}{i}p^{i}(1-p)^{n-i}\nonumber\\
	& = \left\{
			\begin{array}{ccl}
			-n		&;& k=0\\
			0   &;& k>0
			\end{array}
			\right..
\end{align}
Using this result with \eqref{equ:error_var_hard} and \eqref{equ:exit_subchannels} and the fact that for $\epsilon_{av} \to 0$ the reliability of the channel message $D_{ch}$ is small compared to the reliability of the a-priori message $D_{av}$, for regular LDPC codes we get
\begin{align}
\label{equ:stabcoeffs}
\sum_{w=0}^{W}p_{w} \left(\lim_{\eta\to 0} \frac{\partial f_{v}(\eta;d_{v},\epsilon_{ch,w})}{\partial\eta}\right)\nonumber\\
& \hspace{-4cm}= 
\left\{
\begin{array}{ccl}
	1																		&;&	d_{v}=2\\
	2\sum_{w=0}^{W}p_{w}\epsilon_{ch,w}	&;&	d_{v}=3\\
	0																		&;&	d_{v}\geq 4
\end{array}
\right..
\end{align}
The expectation of the error probabilities of all sub-channels equals the error probability of a hard decision channel $\epsilon_{ch} = \sum_{w=0}^{W} p_{w}\epsilon_{ch,w}$. Since every binary input symmetric output channel can be decomposed into BSCs \cite{Land2005,Land2006}, \eqref{equ:stabcoeffs} is also valid for unquantized channel messages by setting $\epsilon_{ch}=Q\left(\frac{\sigma_{ch}}{2}\right)$.
For irregular LDPC codes \eqref{equ:irregular} leads to
\begin{align}
\label{equ:stabvar}
\sum_{i} \left(\lambda_{i} \sum_{w=0}^{W}p_{w} \left(\lim_{\eta\to 0} \frac{\partial f_{v}(\eta;i,\epsilon_{ch,w})}{\partial\eta}\right)\right) = \lambda_{2} + 2\epsilon_{ch}\lambda_{3}.
\end{align}
Combining \eqref{equ:stabchk} and \eqref{equ:stabvar} with \eqref{equ:oneiteration} yields the theorem.
\end{IEEEproof}
We note that this stability condition is the same as for Gallager's original algorithm B (with hard decisions) as derived in \cite[Eq. (11)]{Ardakani2005}, i.e. a finer quantization of the channel messages does not change the stability condition. The stability condition imposes a linear constraint on $\lambda_{2}$ and $\lambda_{3}$ and can hence be incorporated in the linear program \eqref{equ:linprog} for code optimization.

For regular LDPC codes with $d_{v}=2$, the left side of \eqref{equ:stability} is $d_{c}-1$, which cannot be less than one. Therefore, such codes exhibit an error floor with binary message-passing decoding. Regular codes with $d_{v}=3$ satisfy the stability condition if $2\epsilon_{ch}(d_{c}-1)<1$. However, we show in the next section that these codes also exhibit an error floor with binary message-passing decoding.

\subsection{Effect of Cycles}
\label{sec:cycles}
As shown in \cite{Etzion1999}, cycles are unavoidable in the factor graph of finite length LDPC codes. Some cycles lead to error floors because they lead to poor properties of the code itself, e.g. cycles of length $g$ that include only variable nodes of degree two lead to a minimum distance $d_{min}\le\frac{g}{2}$. For binary message-passing decoders, we identify cycles that lead to error floors that are caused by the decoding algorithm and not by the properties of the code.

\begin{theorem}
\label{th:cyclesfloor}
Consider a cycle formed by variable nodes of degree two and three only and assume that the channel messages associated with the nodes in the cycle are in error. If all other incoming messages at the check nodes of the cycle are correct, then the variable nodes forming the cycle cannot be corrected by the binary message-passing decoder.
\end{theorem}
\begin{IEEEproof}
\begin{figure}
  \centering
  \includegraphics[width=4cm]{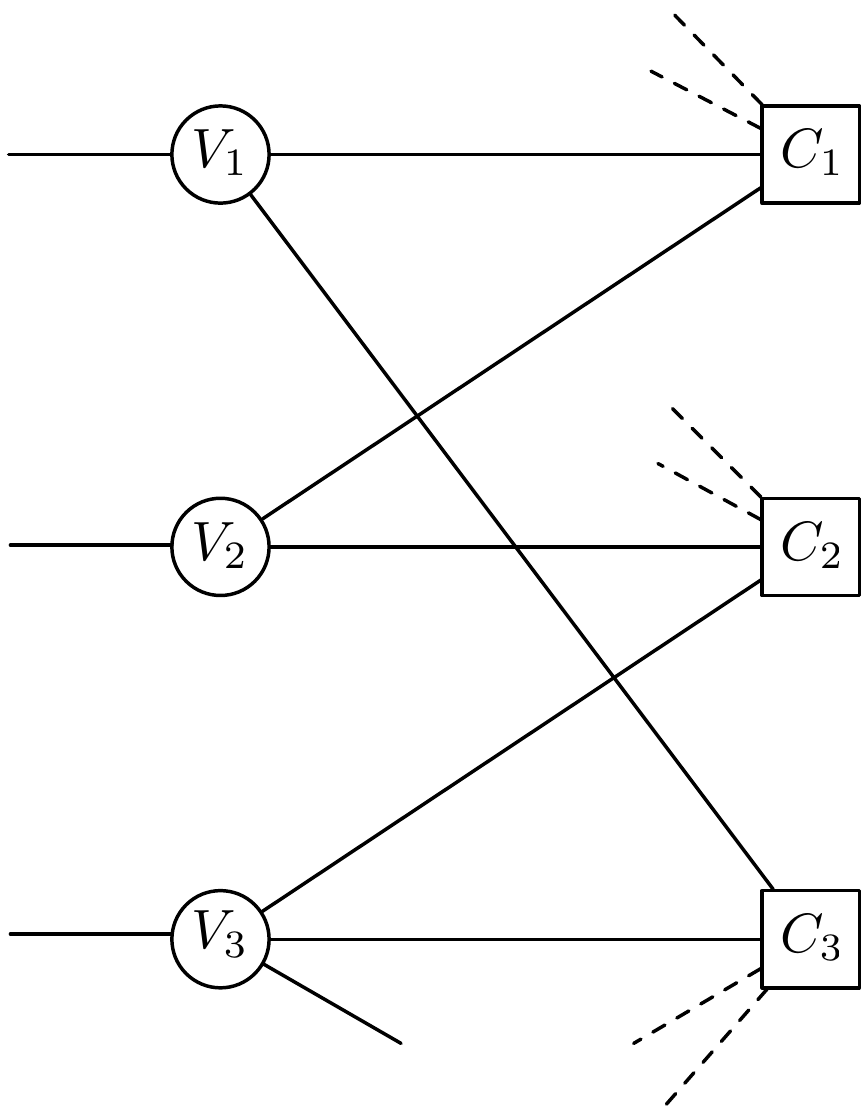}
  \caption{Cycle formed by variable nodes of degree two and three.}
  \label{fig:cycle}
\end{figure}
An example of a cycle of interest is shown in Figure~\ref{fig:cycle}, where the left edges correspond to the channel messages which are all assumed to be in error. In the first iteration the variable nodes send out their received messages. Since every check node in the cycle is connected twice to the set of erroneous variable nodes, the outgoing messages from the check nodes are also in error.

In the following iteration, the outgoing messages at the variable nodes are computed according to \eqref{equ:varoperation}. For degree two variable nodes ($V_{1}$ and $V_{2}$), the extrinsic L-value is the sum of the channel L-value and the L-value of the other incoming message. Since both messages are in error, the outgoing message is also in error. For variable nodes of degree three, one message is not involved in the cycle (as shown for $V_{3}$ in Figure~\ref{fig:cycle}). Even if this message is correct, the extrinsic L-value is the sum of the channel L-value and two L-values from the check nodes with same magnitude but different signs. Therefore, the outgoing message is the sign of the channel L-value which is in error. Since this leads to the same state as after the first iteration, the decoder is not able to correct the errors of the variable nodes in the cycle.
\end{IEEEproof}

If the factor graph contains cycles consisting of variable nodes of degree two and three only, there is a nonzero probability that the involved channel messages are in error leading to a decoding failure. Similar to stopping sets \cite{Richardson2008}, this situation leads to error floors. According to Theorem~\ref{th:cyclesfloor}, to avoid an error floor, the factor graph of an LDPC code must not have such cycles. Such a graph exists if the following condition is satisfied.

\begin{theorem}
\label{th:cyclefree}
A factor graph with no cycles of variable nodes of degree two and three exists if and only if
\begin{equation}
\label{equ:cyclefree}
3\lambda_{2} + 4\lambda_{3} \leq \frac{6}{d_{c}}\frac{(1-R)-\frac{1}{N}}{(1-R)} < \frac{6}{d_{c}}.
\end{equation}
\end{theorem}
\begin{IEEEproof}
Let $\Lambda_{i}$ denote the fraction of variable nodes of degree $i$, i.e. the node perspective of the variable node degree distribution \cite{Richardson2001}. Furthermore, let $N$ and $M$ denote the number of variable and check nodes, respectively. The maximum number of nodes in the subgraph containing only degree two and three variable nodes is
\begin{equation}
\Lambda_{2}N + \Lambda_{3}N + M.
\end{equation}
This subgraph is cycle-free only if the number of edges $E_{t}$ in the sub-graph is at most one less than the number of nodes
\begin{equation}
\label{equ:edgestree}
E_{t} \leq \Lambda_{2}N + \Lambda_{3}N + M - 1.
\end{equation}
Furthermore, such a subgraph exists if \eqref{equ:edgestree} is satisfied. Since $E_{t}=2\Lambda_{2}N + 3\Lambda_{3}N$, the bound \eqref{equ:edgestree} is
\begin{align}
\label{equ:treecond}
\Lambda_{2} + 2\Lambda_{3} 		& \leq \frac{M - 1}{N}.
\end{align}
Using
\begin{equation}
\label{equ:n2e}
\Lambda_{i} = \frac{\frac{\lambda_{i}}{i}} {\sum_{j}\frac{\lambda_{j}}{j}},
\end{equation}
to convert from node perspective $\Lambda_{i}$ to edge perspective $\lambda_{i}$, and expressing the design rate $R$ of the LDPC code as \cite{Richardson2001}
\begin{align}
\label{equ:rate}
R = 1 - \frac{\frac{1}{d_{c}}}{\sum_{j}\frac{\lambda_{j}}{j}} = 1 - \frac{M}{N}
\end{align}
leads to the theorem.
\end{IEEEproof}

An important consequence of Theorems~\ref{th:cyclesfloor} and \ref{th:cyclefree} is that regular LDPC codes with $d_{v}<4$ cannot be decoded without an error floor using binary message-passing decoders. To see this, observe that for regular codes with $d_{v}=2$ or $d_{v}=3$, Theorem~\ref{th:cyclefree} is satisfied only if $d_{c}\leq2$ and $d_{c}\leq1.5$, respectively. Both cases are not possible for codes of positive rate. Therefore, although regular LDPC codes with $d_{v}=3$ are attractive from the point of view of a cycle-free analysis \cite{Ardakani2004a}, they have limited use for binary message-passing. To demonstrate this fact we show simulation results of such a code in Section~\ref{sec:simulations}.

\section{Implementation Aspects}
\label{sec:implementation}

\subsection{Estimation of the A-Priori Channel}
\label{sec:estimation}
In Section~\ref{sec:exit}, we assumed that the variable node knows the parameters of the extrinsic channel for every iteration, i.e. it knows the crossover probabilities of the messages from the check nodes to the variable nodes. Suppose we represent these crossover probabilities as a sequence of numbers indexed by the iteration number. In \cite{Lechner2007a} we showed how this sequence can be predicted in advance from the trajectory of the decoder in the EXIT chart. Another possibility is to determine a sequence of crossover probabilities by simulations. We remark that for finite-length codes, such an approach can lead to better results than an asymptotic prediction from the EXIT chart. In this section, we present an adaptive method which is based on the fraction of unsatisfied check nodes \cite{Yue/Wang:ISTC:2008} and does not require a precomputed sequence of crossover probabilities.

Estimating the error probability based on the extrinsic messages of the check nodes is not possible, because these messages depend on the transmitted symbols which are not known to the decoder. However, the decoder knows that all $d_{c}$ symbols involved in a parity-check equation sum up to zero, and it can therefore determine the extrinsic error probability $\epsilon_{ec}$ from the number of unsatisfied check nodes denoted by $M_{e}$. Denote the fraction of unsatisfied check nodes as
\begin{equation}
\label{equ:syndromerror}
	\epsilon_{s} = \frac{M_{e}}{M}.
\end{equation}
This quantity can be seen as the extrinsic error probability of a check node of degree $d_{c}+1$. Using \eqref{equ:exit_check_inv} with $\epsilon_{s}$ leads to an estimate of the crossover probability at the input of the check node
\begin{equation}
\label{equ:estimate_eac}
  \epsilon_{ac} \approx \frac{1-\left(1-2\epsilon_{s}\right)^{\frac{1}{d_{c}}}}{2}.
\end{equation}
The crossover probability probability at the output of the check node follows from \eqref{equ:exit_check} and \eqref{equ:estimate_eac}
\begin{equation}
  \epsilon_{ec} = \frac{1-\left(1-2\epsilon_{ac}\right)^{d_{c}-1}}{2} \approx \frac{1-\left(1-2\epsilon_{s}\right)^{\frac{d_{c}-1}{d_{c}}}}{2}.
\end{equation}
This estimate is then used in \eqref{equ:reliability} to compute the reliability of the extrinsic channel.

\subsection{Majority Decision Rules}
\label{sec:majority}
The concept of converting all incoming messages to L-values, perform decoding in the L-value domain and sending the hard decision from variable to check nodes is useful for analyzing binary message-passing decoders. In practice, however, these operations are often replaced by majority decision rules due to reasons of complexity. Consider a variable node of degree $d_v$. Depending on $\epsilon_{av}$, a minimum number of $t$ messages (see \eqref{equ:tdefinition}) from the check nodes have to disagree with the channel messages, in order to change the outgoing message of that variable node. This allows the derivation of a majority decision rule that is parametrized by $\epsilon_{av}$. Since $\epsilon_{av}=\epsilon_{ec}$ we can use the fraction of unsatisfied check nodes to adapt the majority decision rule over the iterations. For channels with larger output alphabet than binary, a majority decision rule has to be defined for every sub-channel (see \eqref{eq:pk}-\eqref{eq:Dch}).

\section{Numerical Results}
\label{sec:simulations}
\begin{figure}
  \centering
  \includegraphics[width=\figwidth]{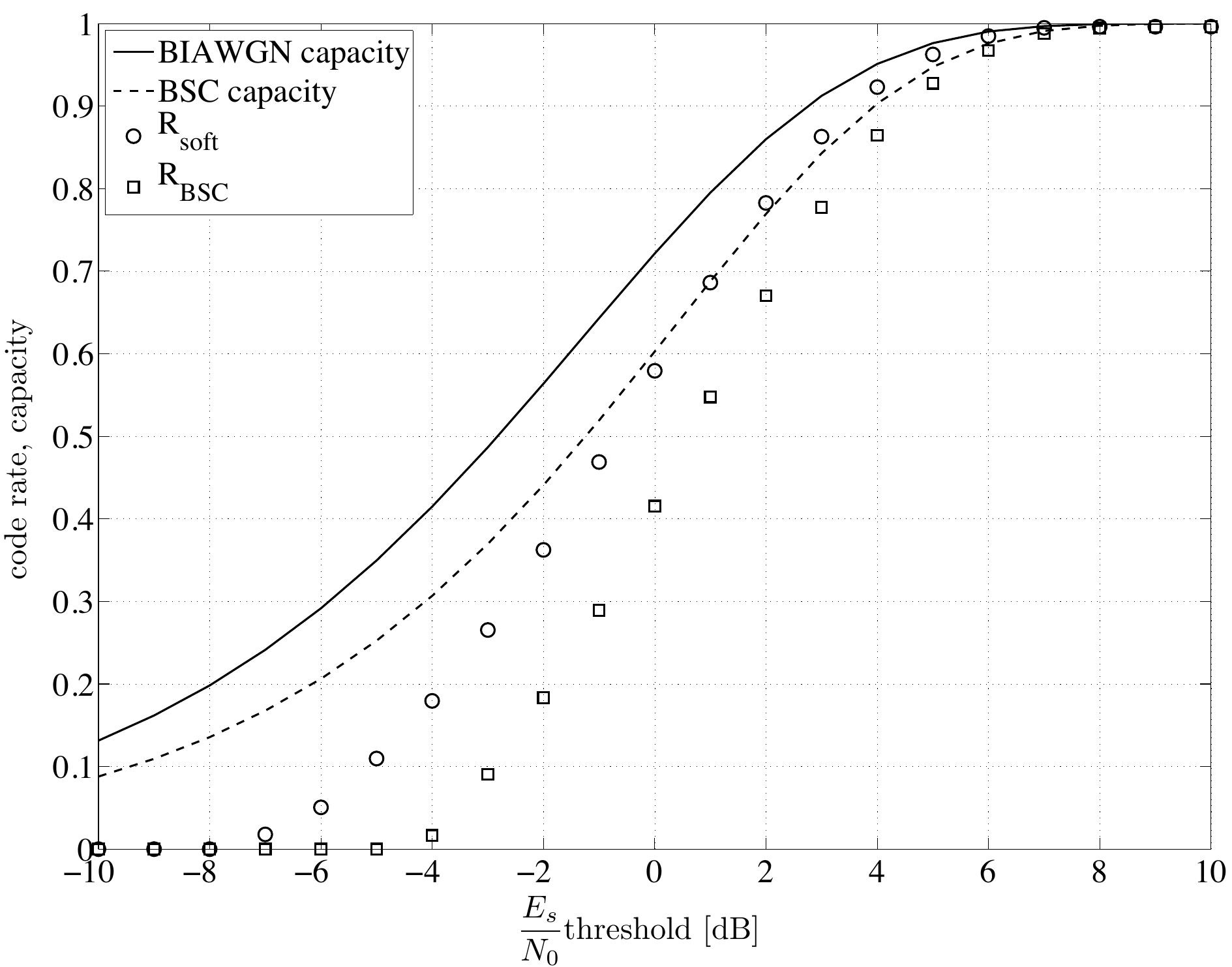}
  \caption{Thresholds of optimized codes for soft channel information ($R_{\mbox{\scriptsize{soft}}}$) and hard decision channel ($R_{\mbox{\scriptsize{BSC}}}$).}
  \label{fig:thresholds}
\end{figure}

Using linear programming, we optimized codes using \eqref{equ:linprog} with \eqref{equ:stability} and \eqref{equ:cyclefree} and compared them with the capacity of the BIAWGN and BSC. For the optimization we set the maximum variable node degree to $d_{v,max}=100$ and performed the optimization for check node degrees in the range between 2 and 1000. The decoding thresholds of these codes in terms of the required signal to noise ratio are shown in Figure~\ref{fig:thresholds}. Note that cycles of the kind described in Theorem~\ref{th:cyclesfloor} could occur if a random interleaver was used. However, we constructed our codes using the progressive edge-growth (PEG) algorithm \cite{Hu2005,PEGSoftware} which ensures that such cycles do not exist.

Observe that the gap to capacity decreases as the rate increases. This makes binary message-passing decoders attractive for applications which require a high code rate. For a rate of 0.9, the best code using soft channel information is within approximately 0.5 dB of capacity. Note that the soft-decision and hard-decision channel EXIT curves serve as upper and lower bounds, respectively, for all quantization schemes.

\begin{figure}[ht]
  \centering
  \includegraphics[width=\figwidth]{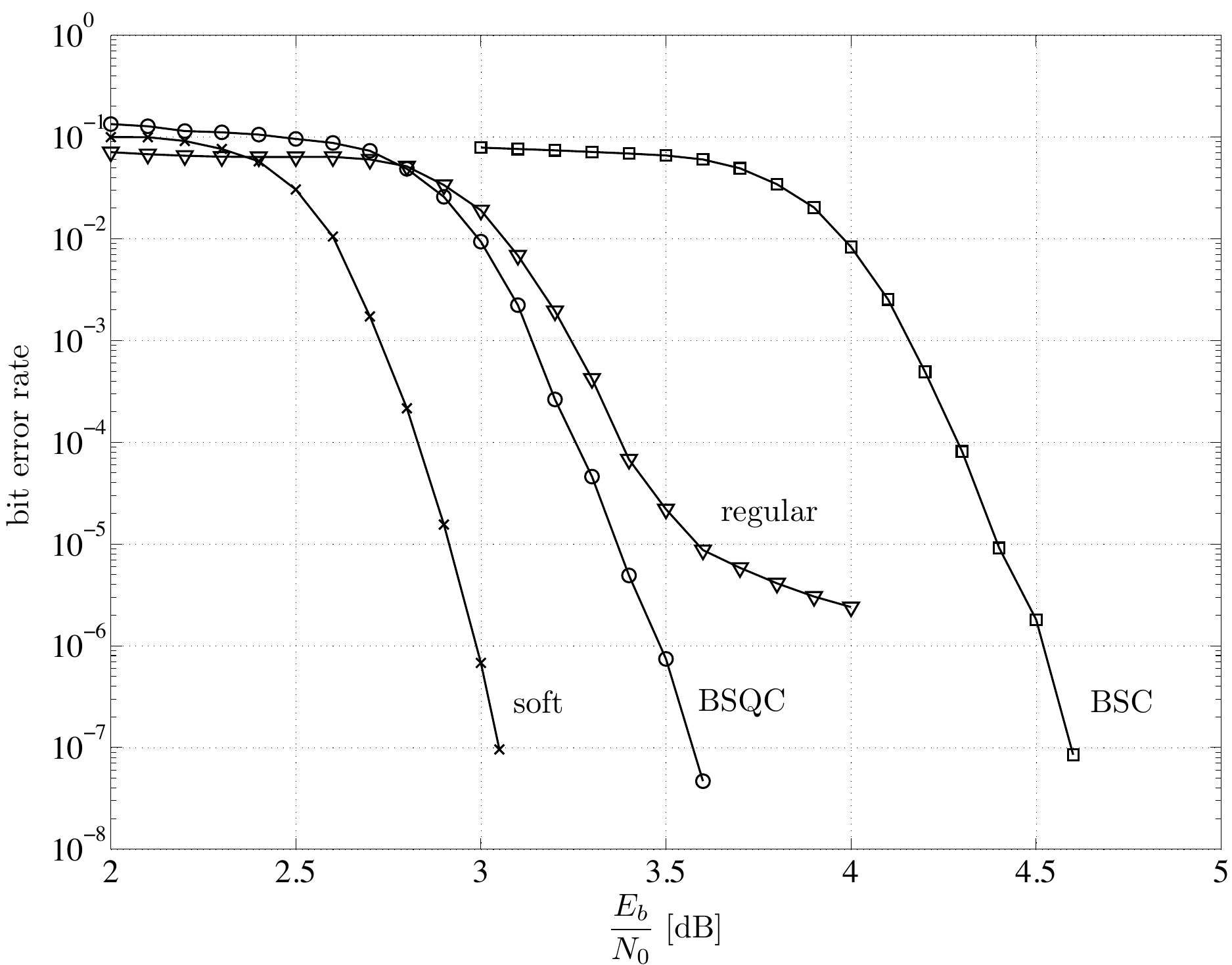}
  \caption{Bit error rate simulations for optimized and regular code of rate $R=0.5$.}
  \label{fig:ber}
\end{figure}

We performed bit error rate simulations for codes of rate $R=0.5$ for the BSC, BSQC and the soft information channel. The thresholds of these codes using the associated quantization schemes are shown in Table~\ref{tab:thresholds} where $\zeta_{1}$ is the quantization interval (see Section~\ref{sec:quantexamples}) that leads to the best threshold. The bit error rate simulation results are shown in Figure~\ref{fig:ber} using codes of length $N=10^4$. Also shown in this figure are the results for a regular LDPC code with variable node degree $d_{v}=3$ and check node degree $d_{c}=6$ transmitted over a channel with soft outputs. As expected from Section~\ref{sec:cycles} this code shows an error-floor.

\begin{table}
\centering
\caption{Thresholds for code of rate 0.5 and the LDPC code used in \cite[Sec. I.6]{ITU-T2004}.}
\label{tab:thresholds}
\begin{tabular}{cccccccc}
\hline
\multicolumn{1}{c}{}			& \multicolumn{3}{c}{Rate 0.5} & & \multicolumn{3}{c}{Rate 0.9375}\\
\cline{2-4}\cline{6-8}
 							& BSC  		& BSQC 		& Soft		& & BSC			& BSQC			& Soft\\
\hline
$E_b/N_0$[dB] 	& $3.67$ 	& $2.69$ 	& $2.30$	& &	$6.08$		& $5.12$		&	$5.02$\\
$d_{c}$ 				&	$15$			&	$14$			& $12$		& &	$112$		& $112$		& $112$\\
$\zeta_{1}$		&					&	$1.95$		&				& &					& $2.34$		&\\
\hline
\end{tabular}
\end{table}

The system with hard channel decisions (BSC) corresponds to the Gallager B algorithm. Observe that by adding one more bit for the channel messages and quantizing them according to a BSQC, the performance of this algorithm can be improved by 1.0~dB with only a small increase in decoding complexity. A finer quantization of the channel messages will not result in a significant gain, since the gap to the unquantized system is only approximately 0.5~dB. Remark that predicting the sequence of crossover probabilities by using the trajectory in the EXIT chart \cite{Lechner2007a} gives a similar performance.

\begin{figure}[ht]
  \centering
  \includegraphics[width=\figwidth]{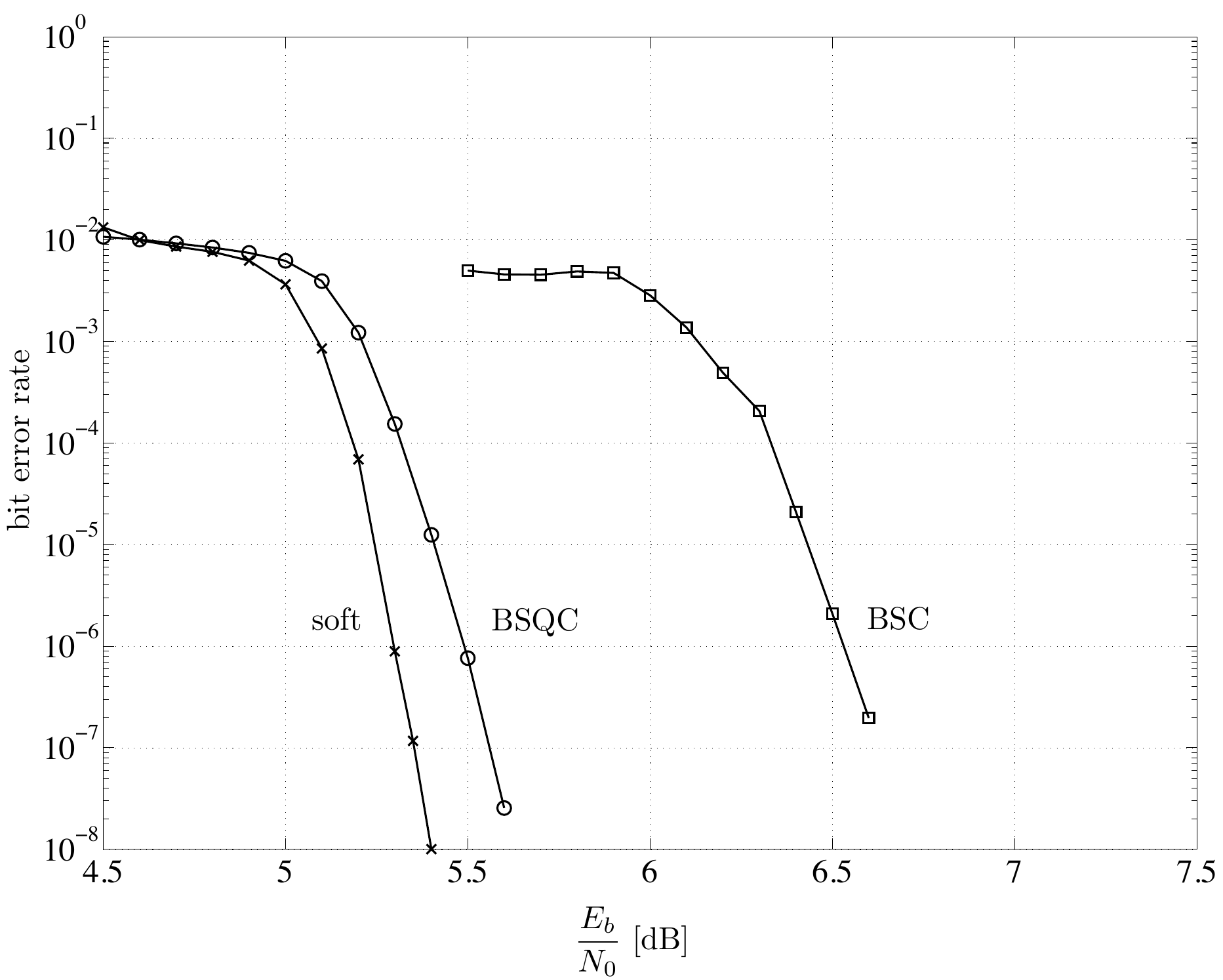}
  \caption{Bit error rate simulations for the LDPC code of rate $R=0.9375$ used in \cite[Sec. I.6]{ITU-T2004}.}
  \label{fig:ber2}
\end{figure}

To demonstrate the performance of binary message-passing decoders for high code rates we perform bit error rate simulations of a code used in \cite[Section I.6]{ITU-T2004}. This code of rate $R=0.9375$ is a regular LDPC code with $d_v=7$, $d_c=112$ and is used for an optical transmission system for high data rates up to 40 Gbps. The slopes which define the code (for details see \cite[Section I.6.2]{ITU-T2004}) are not defined in the standard and have been chosen to be the first seven prime numbers, i.e., $2, 3, 5, 7, 11, 13, 17$. In \cite{ITU-T2004} it is assumed that the decoder observes hard decisions from the channel. Using our analysis, the thresholds for channels with larger output alphabets are shown in Table~\ref{tab:thresholds} assuming that the unquantized channel can be modeled as a BIAWGN channel. The corresponding bit error rate simulations are shown in Figure~\ref{fig:ber2}. Increasing the number of quantization levels of the channel messages by one bit leads to an improvement of approximately 1.0~dB.

\section{Conclusions}
\label{sec:conclusions}
We analyzed binary message-passing decoders using EXIT charts. For channels which deliver hard decisions, this analysis led to an algorithm that is equivalent to Gallager's decoding algorithm B. We extended these results to channels with larger output alphabets including channels providing soft information. We found that increasing the channel output alphabet size by only one bit leads to a significant lowering of the decoding threshold. We described why the mixing property of EXIT functions does not apply to binary message-passing algorithms if one uses mutual information, and presented a modified mixing method based on error probabilities in order to optimize codes. Finally, we showed that degree two and three variable nodes forming a cycle cannot be corrected and we derived a condition for the variable node distribution that guarantees the existence of a cycle-free subgraph for these nodes.
\newpage
\bibliographystyle{IEEEtran}
\bibliography{bibnames,bmp-journal}

\end{document}